# Self-injection-locked dark-soliton Kerr optical frequency combs with high even-order normal dispersion

Lei Shi[1], Rujiang Li[2], Shilong Liu[3], Dumitru Mihalache[4], Boris A. Malomed[5,6] and Pengfei Li[1,*]

[1]*Department of Physics, Taiyuan Normal University, Jinzhong, 030619, China*
[2]*National Key Laboratory of Radar Detection and Sensing, School of Electronic Engineering, Xidian University, Xi'an 710071, China*
[3]*femtoQ Lab, Department of Engineering Physics, Polytechnique Montréal, Montréal QC H3T 1J4, Canada*
[4]*Horia Hulubei National Institute of Physics and Nuclear Engineering, Magurele, Bucharest RO-077125, Romania*
[5]*Department of Physical Electronics, School of Electrical Engineering, Faculty of Engineering, and Center for Light-Matter Interaction, Tel Aviv University, Tel Aviv 69978, Israel*
[6]*Instituto de Alta Investigación, Universidad de Tarapacá, Casilla 7D, Arica, Chile*
**Corresponding author: lpf281888@gmail.com*

**We numerically investigate dissipative dark-soliton (DDS) microcombs under the action of pure high even-order normal (HEON) dispersion in a microresonator with the self-injection locking (SIL). Pure 4th, 6th, 10th, and 14th-order normal dispersion can support DDSs formed by interlocked switching waves, directly excited through a SIL-based turnkey scheme without detuning sweeps. Compared to the pure 4th-order case, the switching-wave tail oscillation period increases and its decay rate decreases with the increase of the dispersion order. Although the overall comb bandwidth narrows, the number of non-pump comb lines above a common absolute power threshold increases. Furthermore, the feedback phase controls the DDS duty cycle and intracavity comb-line power distribution, while the increase of the pump power improves the estimated pump-to-comb output conversion efficiency.**

The generation of coherent light sources based on optical Kerr frequency combs (OKFCs) in optical cavities has found applications in many fields, including optical communications, high-precision metrology, parallel optical computing, and spectroscopy, owing to the broad spectrum, high coherence, and potential for on-chip miniaturization offered by OKFCs [1–4]. Coherent, mode-locked OKFCs can be generated through dissipative solitons, which are maintained by the double balance between dispersion and Kerr nonlinearity, as well as between loss and gain [5–8]. Characteristics of dissipative solitons are determined by the dispersion order and sign in the cavity. A new approach to the design of high-performance OKFCs is based on the use of even-order dispersion [9, 10]. Compared to the systems with the second-order dispersion, the ones using pure high even-order anomalous dispersion not only support dissipative bright solitons with higher energy [11, 12], but also flatten the spectral profile and facilitate the formation of perfect soliton crystals [13].

Previous studies have indicated that stable dissipative dark pulses, as well as platicons (flat-top pulses), can be formed in optical cavities under normal dispersion [14–16]. Platicons were first proposed in Ref. [16], and their dynamics under third-order dispersion were subsequently studied in Ref. [17]. These dissipative structures can exhibit high pump-to-comb conversion efficiency [18–22]. Recent research has established that fourth-order dispersion (FOD) systems support dissipative dark solitons (DDSs) with oscillatory tails and relatively flat comb spectra [10, 23, 24]. However, the dependence of the switching-wave dynamics on the dispersion order remains unclear, in particular the tail oscillation period, decay rate, and comb-line power distribution. Furthermore, it is necessary to examine the possibility of directly exciting and controlling these states through self-injection locking (SIL).

In this Letter, we focus on dissipative dark-soliton OKFCs under pure high even-order normal (HEON) dispersion. By numerically solving the Lugiato-Lefever equation (LLE) with SIL feedback, we investigate the temporal and spectral characteristics of DDSs. It is found that pure quartic-, sextic-, decic-, and fourteenth-order normal dispersion can support DDSs formed by interlocked switching waves, with direct SIL-based turnkey excitation. As the dispersion order increases, the tail oscillation period increases and its decay rate decreases, while the number of non-pump comb lines above a common power threshold increases despite narrower spectra. Additionally, based on the turnkey scheme, we observe that the feedback phase can effectively adjust the duty cycle of dark solitons and intracavity power, while the pump power improves the estimated pump-to-comb output conversion efficiency.

It is challenging to directly generate OKFCs by means of conventional methods, such as frequency sweeping under normal-dispersion conditions. This is explained by the difficulty in initiating modulation instability (MI), which is readily achievable in the opposite case of anomalous dispersion [25–27]. Recently, a turnkey scheme integrating the SIL effect has been reported for microcombs under anomalous and normal dispersion [28, 29]. It eliminates the need for complex frequency tuning under suitable operating conditions. The dynamics of the turnkey DDS is governed by the system of coupled LLEs, along with an algebraic equation modeling the injection-locking dynamics [30, 31]:

$$\frac{\partial A}{\partial t} = -\left(1+i\delta\right)A + \sum_{k\ge 2}\frac{i^{k+1}d_k}{k!}\frac{\partial^k A}{\partial\theta^k} + i\left(\left|A\right|^2 + 2\overline{\left|B\right|^2}\right)A + i\eta^*\overline{B} + F, \tag{1}$$

$$\frac{\partial B}{\partial t} = -\left(1+i\delta\right)B + i\left(\left|B\right|^2 + 2\overline{\left|A\right|^2}\right)B + i\eta\overline{A}, \qquad (2)$$

$$\delta = \delta_L + K\,\mathrm{Im}\left(e^{i\Phi}\overline{B}/F\right), \qquad (3)$$

where the normalized forward and backward intracavity fields are $A = \sqrt{\gamma L/\alpha}E_A$ and $B = \sqrt{\gamma L/\alpha}E_B$, the pump is $F = \sqrt{\gamma L\kappa/\alpha^3}E_{in}$, and $d_k = L\beta_k\left(2\pi/t_R\right)^k/\alpha$ is the scaled HEON dispersion coefficient. Here, $\gamma, L, \alpha, \beta_k, t_R, \kappa$ and $E_{in}$ represent the nonlinear coefficient, the cavity length, the total loss, the dispersion parameter, the round-trip time, the coupling coefficient and the pump amplitude, respectively. Further, $t$, $\theta$, $\delta$, $\eta$, $\delta_L$, $K$, and $\Phi$ stand for the evolution time, azimuthal-angle coordinate along the cavity ring, detuning, backscattering coefficient, initial detuning, locking strength, and feedback phase, respectively. Further, $\overline{|A,B|^2} = \int_0^{2\pi}|A,B|^2\,d\theta/2\pi$ are the average intracavity powers, while $\overline{A,B} = \int_0^{2\pi}A,B\,d\theta/2\pi$ are the average intracavity fields. The backscattered light in the cavity provides the feedback to the laser, which controls the effective detuning, while the combined effect of the nonlinearity and dispersion gives rise to the OKFC formation in the cavity. Here, we mainly consider the cases of dark solitons under the action of HEON dispersion, where $k$ takes the values of 4, 6, 10 or 14. The normalization was performed with physically relevant parameters: $L = 3.14\,\text{mm}, E_{in} = \sqrt{100\,\text{mW}}, t_R = 36.153\,\text{ps}, \delta_L = 0, \Phi = 1.9\pi$, $\gamma = 0.5056\ \text{W}^{-1}\text{m}^{-1}$. The total cavity loss and coupling coefficients are $\alpha = \kappa = 0.007235$, satisfying the critical coupling condition, with a normalized backscattering coefficient $\eta = 0.01382$ and locking strength $K = 2000$. For comparison across dispersion orders, the dispersion coefficients are scaled according to $\beta_k = \frac{\beta_2 k!}{2\Delta\omega^{k-2}}$, where $\beta_2 = 90\ \text{ps}^2/\text{km}$ and $\Delta\omega = 1.6 \times 10^{12}$ Hz [13, 32]. The simulations use 1024 grid points per round trip, corresponding to a fast-time resolution of 35.31 fs and a frequency resolution of 27.66 GHz. The forward intracavity field is initialized with weak random noise.

Figure 1 illustrates temporal and spectral characteristics of mode-locked dark solitons with high-even-order normal dispersion, obtained by solving Eqs. (1) to (3) using the split-step Fourier method. Figure 1(a) shows the steady-state temporal envelopes for $k$ = 4, 6, 10, and 14. These states consist of two switching waves connecting the upper and lower continuous-wave (CW) steady-state branches of the intracavity bistability curve [33]. Figures 1(b–d) reveal stronger oscillations near the lower CW state, consistent with the different Kerr-induced phase shifts and linearized responses of the two CW states [10, 33]. Lower dispersion orders exhibit steeper switching-wave fronts. Figures 1(e, f) quantify the oscillation periods and fitted decay rates of the switching-wave tails. As $k$ increases from 4 to 14, the oscillation period increases from 0.0869π to 0.1650π, while the decay rate decreases from 8.331 to 1.351 $\pi^{-1}$, indicating longer-period oscillations and weaker damping. Figures 1(g–j) demonstrate that the increase of the dispersion order leads to a progressive contraction of the spectral profile. Here, $\mu = m - m_p$ denotes the relative spectral mode index, where $m_p$ is the pumped-mode number and $\mu = 0$ corresponds to the pumped mode. Spectral-edge peaks appear for $k$ = 6, 10, and 14, but further analysis is required to clarify their origin. At a common absolute intracavity mode-power threshold of 9 dBm, the numbers of non-pump comb lines are 12, 16, 20, and 22 for $k$ = 4, 6, 10, and 14, respectively.

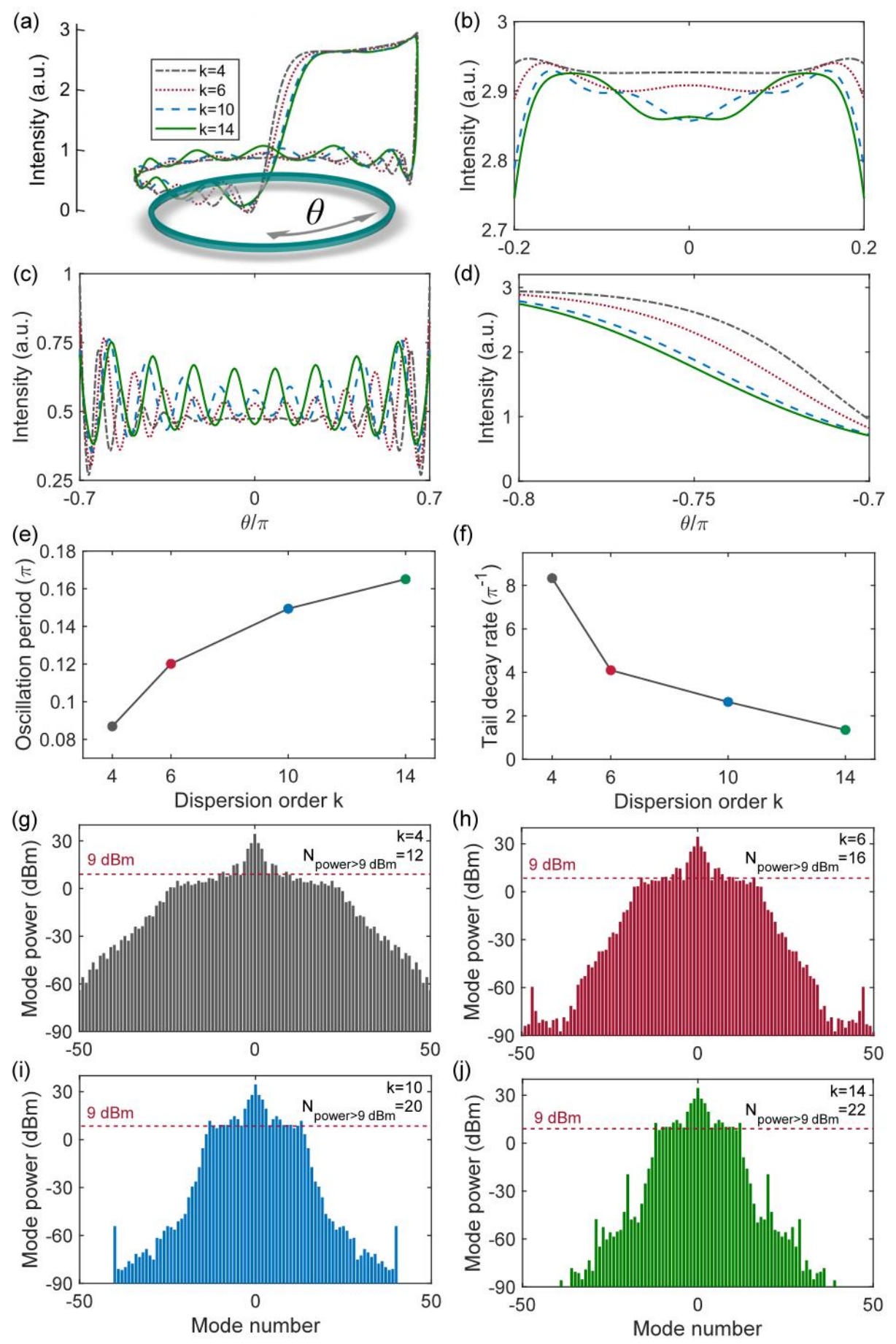

Fig. 1. Temporal and spectral characteristics of dark solitons under the action of pure HEON dispersion. (a) Intracavity intensity profiles mapped along the azimuthal angle θ of the microresonator for dispersion orders k = 4, 6, 10, and 14. (b-d) Magnified views of the upper CW branch, lower CW branch, and switching-wave front, respectively. (e, f) The oscillation period and fitted decay rate of the switching-wave tails versus the dispersion order, respectively. (g-j) The corresponding absolute intracavity mode-power spectra for k = 4, 6, 10, and 14, respectively. Red dashed lines indicate the common 9 dBm threshold, with N denoting the number of non-pump comb lines.

We employ the turnkey scheme to effectively excite DDSs. The intracavity state evolves to a mode-locked state without frequency scanning. Figure 2 illustrates the evolution of normalized average intracavity power versus normalized detuning under different HEON dispersions (e.g., sextic-, decic-, and fourteenth-order dispersion). The simulation parameters are identical to those in Fig. 1. The black dashed curve shows the Kerr-tilted CW response as a reference for the SIL evolution trajectories. Figure 2(a) shows the overall evolution for three cases. Starting with weak random noise in the forward field, a zero backward field, and $\delta_L = 0$ (hence $\delta = 0$ initially), the system first evolves toward the red-detuned region. Due to self-phase and cross-phase modulation, the state is pulled toward the turnkey operating point [29, 34]. Consequently, the detuning undergoes a blueshift, accompanied by an increase in intracavity power. Ultimately, the three cases reach effective red-detuned DDS states. Figure 2(b) resolves the different terminal states for $k$ = 6, 10, and 14. As $k$ increases, the final detuning and normalized average intracavity power decrease. As shown in Fig. 2(c), the dark pulse remains stable following a random complex-field perturbation near 1085 ns, throughout the simulated 100,000 round trips (≈ 3615 ns), demonstrating long-term robustness at this operating point. For $k$ = 6, 20 independent startups initialized with weak random noise under identical SIL parameters yield terminal profiles containing 1–5 switching-wave pairs with occurrence percentages of 30%, 40%, 15%, 10%, and 5%, respectively [Fig. 2(d)]. Thus, SIL enables direct access to the single-pair DDS without a prescribed pulse seed or detuning sweep, but does not uniquely select this state.

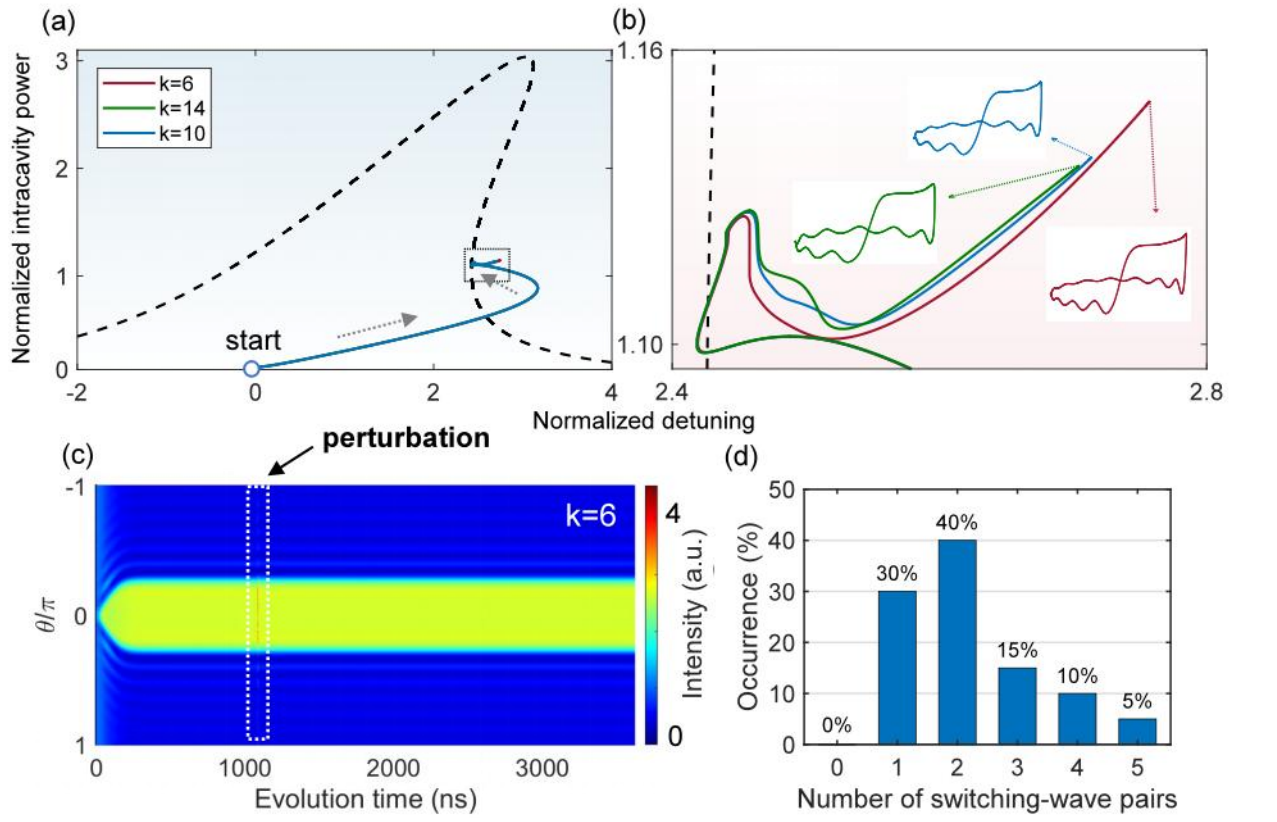

Fig. 2. SIL-based turnkey dynamics and switching-wave state statistics. (a) The evolution of the normalized average intracavity power versus the normalized detuning. Arrows indicate the evolution direction, and the black dashed curve represents the reference CW Kerr response calculated at 100 mW (F = 1.742). (b) A magnified view of the dashed box in (a), where red, blue, and green curves correspond to dispersion orders $k$ = 6, 10, and 14, respectively. Insets depict the corresponding steady-state DDSs in the cavity. (c) The intracavity intensity evolution for $k$ = 6 at Φ = 1.9π and the pump power of 100 mW. The white dashed box marks the perturbation. (d) Occurrence percentages of 0–5 switching-wave pairs from 20 independent SIL startups initialized with weak random noise under the same parameters as (c), each evolved for 100,000 round trips.

With all other parameters held constant as in Fig. 1, Fig. 3(a) depicts the evolution of normalized average intracavity power versus normalized detuning under varying feedback phases (taking k = 6 as a representative case). As Φ increases from 1.70π to 2.00π, the terminal intracavity power decreases while the final detuning remains nearly unchanged. This behavior demonstrates that the feedback phase effectively reconfigures the accessible evolutionary paths, thereby regulating the SIL operating state [29]. Figure 3(b) illustrates the dependence of the dark-soliton duty cycle on the feedback phase. The duty cycle decreases approximately linearly with increasing feedback phase for $k$ = 6, 10, and 14, falling from 43.36% to 29.49% for $k$ = 6. The corresponding intensity profiles [Fig. 3(c)] show a narrowing high-intensity domain with nearly unchanged peak intensity. As Φ increases, Fig. 3(d) shows that the overall spectral envelope and span remain nearly unchanged despite the decreasing total intracavity power. Meanwhile, the power fraction of each $\mu$ = ±1 mode decreases from 16.64% to 10.34%, whereas that of each $\mu$ = ±2 mode increases from 3.53% to 6.38%. These results demonstrate that the feedback phase provides control over the dark-soliton temporal structure and intracavity comb-line power distribution.

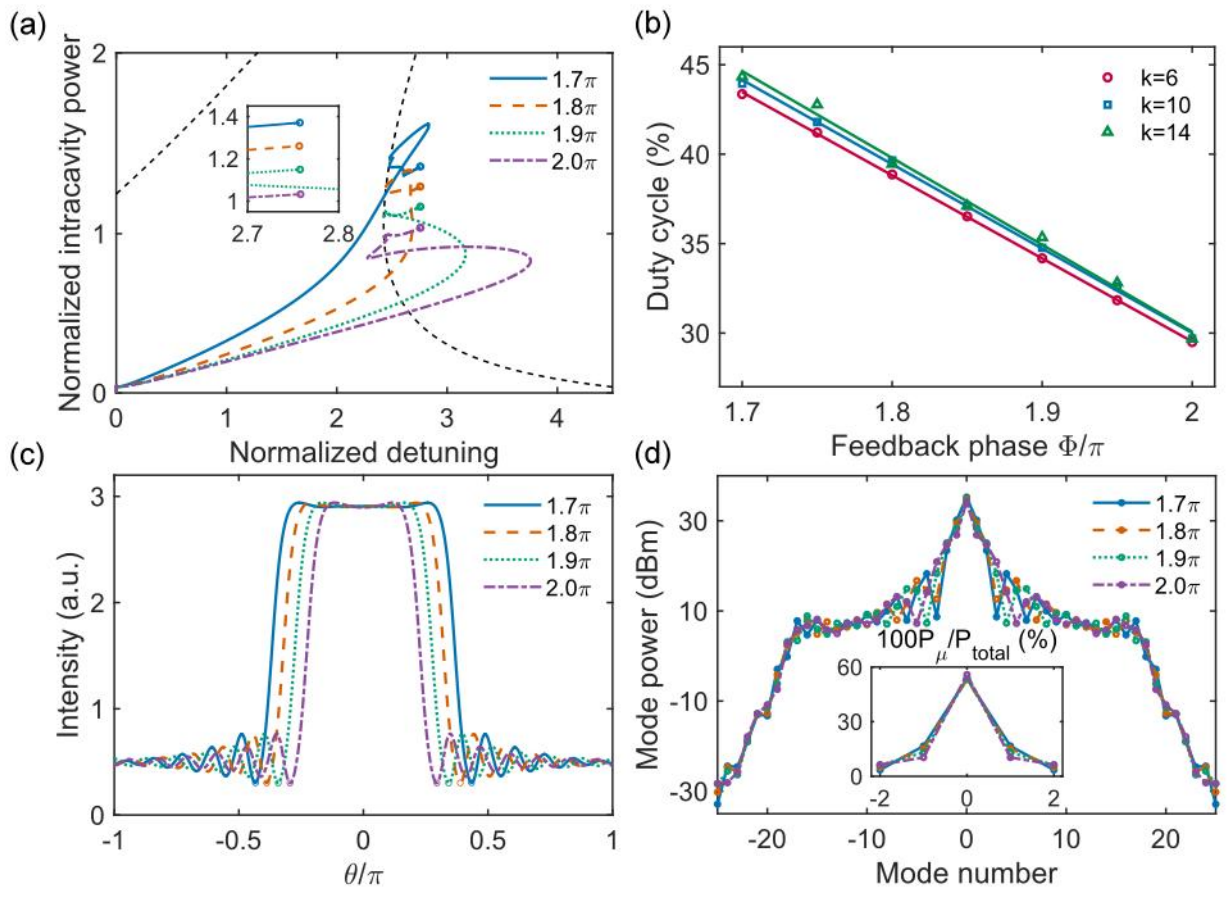

Fig. 3. The effect of feedback phase on cavity dynamics and dark-soliton properties. (a) The evolution of normalized average intracavity power and normalized detuning for different feedback phases Φ. For k = 6, Φ takes values of 1.7π, 1.8π, 1.9π, and 2.0π. The inset magnifies the terminal states. (b) The dependence of dark-soliton duty cycle on the feedback phase for $k$ = 6, 10, and 14 at seven phase values. Symbols denote numerical data and solid lines are linear fits. (c) The corresponding intracavity intensity profiles and (d) the absolute intracavity mode-power spectra for the four phases in (a). The inset in (d) shows the central-mode power fractions, $100P_{\mu}/P_{\text{total}}$ ($\mu$ = −2 to 2), where $P_{\text{total}}$ includes all modes.

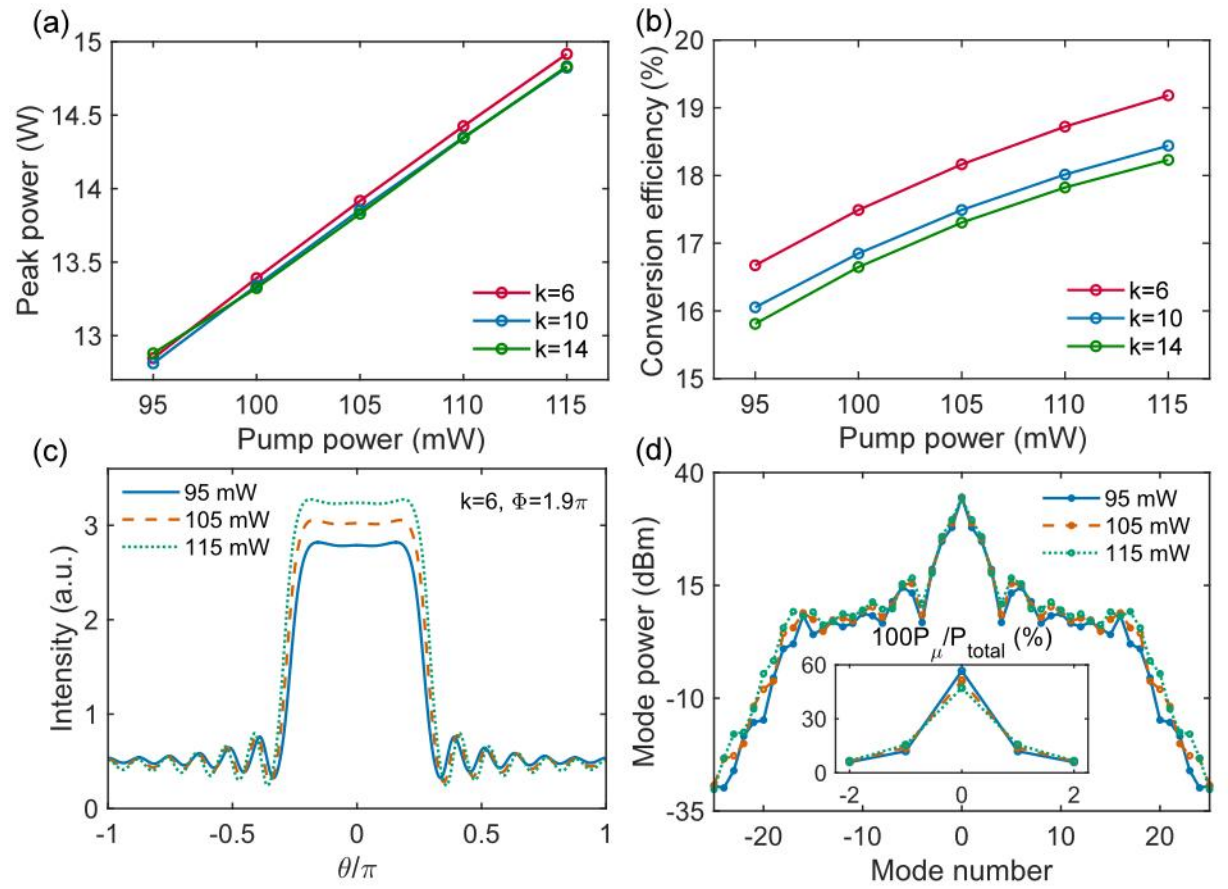

Fig. 4. Scaling of the dark-soliton peak power and pump-to-comb output conversion efficiency with the pump power. (a) The peak intracavity power and (b) the conversion efficiency as functions of the pump power for different dispersion orders. The red, blue, and green curves correspond to $k$ = 6, 10, and 14, respectively. (c) Intracavity intensity profiles and (d) the corresponding absolute intracavity mode-power spectra for $k$ = 6 at pump powers of 95, 105, and 115 mW. The feedback phase is Φ = 1.9π. The inset in (d) shows the central-mode power fractions, $100P_{\mu}/P_{\text{total}}$ ($\mu$ = −2 to +2), where $P_{\text{total}}$ includes all modes.

Figure 4 illustrates the variation in the peak intracavity power and pump-to-comb conversion efficiency of dark solitons, following the variation of the pump power, while other parameters remain constant. The feedback phase is Φ = 1.9π. As the pump power gradually increases, the peak intracavity power increases for all three dispersion orders [Fig. 4(a)]. Figure 4(b) quantifies the estimated output conversion efficiency, $\eta_{\text{conv}} = \kappa P_{\text{comb}}/P_{\text{in}}$, where $P_{\text{comb}}$ is the total intracavity non-pump comb power, assuming the mode-independent coupling [18]. For $k$ = 6, $\eta_{\text{conv}}$ increases from 16.67% to 19.18% as the pump power rises from 95 to 115 mW, with similar trends for $k$ = 10 and 14. The corresponding intensity profiles [Fig. 4(c)] show an increasing peak intensity and slight broadening of the high-intensity domain. The spectra [Fig. 4(d)] retain similar overall envelopes but exhibit nonuniform mode-power redistribution. For $k$ = 6, the total non-pump comb power increases from 2.19 to 3.05 W, while the pumped-mode power decreases from 2.86 to 2.72 W. This redistribution supports the enhanced pump-to-comb conversion shown in Fig. 4(b).

In conclusion, we have investigated dissipative dark solitons in optical Kerr frequency combs with pure HEON (high-even-order-normal) dispersion. Numerically solving the respective Lugiato-Lefever equation, we have explored the temporal-spectral characteristics and parameter dependence of dark solitons. As the dispersion order increases, the oscillation period of the switching-wave's tail increases too, while its fitted decay rate decreases. Although the overall spectral envelope narrows, the number of non-pump comb lines above a common absolute power threshold increases. Our results demonstrate that adjusting the feedback phase makes it possible to regulate intracavity power levels and the dark-soliton's duty cycle, while the increase of the pump power improves the estimated pump-to-comb output conversion efficiency. SIL (self-injection locking) also enables direct turnkey access to DDS states without detuning sweeps.